# Improvement of the perovskite photodiodes performance via advanced interface engineering with polymer dielectric

A.P. Morozov[1§], L.O. Luchnikov[1§], S. Yu. Yurchuk[2], A.R. Ishteev[1], P.A. Gostishchev[1], S.I. Didenko[2], N.S. Saratovsky[3], S.S. Kozlov[4], D.S. Muratov[5*], Yu. N. Luponosov[3*] and D.S. Saranin[1*]

[1]LASE – Laboratory of Advanced Solar Energy, NUST MISiS, 119049 Moscow, Russia
[2]Department of semiconductor electronics and device physics, NUST MISiS, 119049 Moscow, Russia
[3]Enikolopov Institute of Synthetic Polymeric Materials of the Russian Academy of Sciences (ISPM RAS), Profsoyuznaya St. 70, Moscow, 117393, Russia
[4]Laboratory of Solar Photoconverters, Emanuel Institute of Biochemical Physics, Russian Academy of Sciences, 119334 Moscow, Russia
[5]Department of Chemistry, University of Turin, 10125, Turin, Italy

§: The authors contributed equally to this work

**Corresponding authors:** Dr. Danila S. Saranin saranin.ds@misis.ru, Dr. Dmitry S. Muratov dmitry.muratov@unito.it

**Abstract**

Halide perovskite-based photodiodes are promising for efficient detection across a broad spectral range. Perovskite absorber thin-films have a microcrystalline morphology, characterized by a high density of surface states and defects at inter-grain interfaces. In this work, we used dielectric/ferroelectric poly(vinylidene-fluoride-trifluoroethylene) (P(VDF-TrFE)) to modify the bulk interfaces and electron transport junction in p-i-n perovskite photodiodes. Our complex work demonstrates that interface engineering with P(VDF-TrFE) induces significant Fermi level pinning, reducing from 4.85 eV for intrinsic perovskite to 4.28 eV for the configuration with dielectric interlayers. Modifying the interfaces in the devices resulted in an increase in the key characteristics of photodiodes compared to pristine devices. The integration of P(VDF-TrFE) into the perovskite film didn't affect the morphology and crystal structure, but significantly changed the charge transport and device performance. IV curve analysis and 2-diode model calculations showed enhanced shunt properties, a decreased non-ideality factor, and reduced saturation dark current. We have shown that the complex introduction of P(VDF-TrFE) into the absorber's bulk and on its surface is essential to reduce the impact of the trapping processes. For P(VDF-TrFE) containing devices, we increased the specific detectivity from $10^{11}$ to ~$10^{12}$ Jones, expanded the linear dynamic range up to 100 dB, and reduced the equivalent noise power to $10^{-13}$ W·Hz$^{-1/2}$. Reducing non-radiative recombination contributions significantly enhanced device performance, improving rise/fall times from 6.3/10.9 μs to 4.6/6.5 μs, and achieved photo-response dynamics competitive with state-of-the-art analogs. The cut-off frequency (3dB) increased from 64.8 kHz to 74.8 kHz following the introduction of the dielectric. We also demonstrated long-term stabilization of PPD performance under heat-stress. These results provide new insights into the use of organic dielectrics and an improved understanding of trap-states/ion defect compensation for detectors based on perovskite heterostructures.

Halide perovskite (**HPs**) based photodiodes (**PD**s) have emerged as a promising technology in the field of thin-film optoelectronics for a wide array of applications, including medical imaging[1], high-resolution sensing[2], and optical communication[2]. Perovskite refers to a class of materials with the general chemical formula $ABX_3$[3], where the A-cation is an organic cation ($CH(NH_2)_2^+$ - $FA^+$) or inorganic cesium ($Cs^+$); the B-cation is typically lead ($Pb^{2+}$), and the X-anion is iodine (I), bromine ($Br^-$), or chlorine ($Cl^-$). These materials exhibit unique combination of semiconductor properties for efficient photoelectric conversion, including strong optical absorption ($>10^4$ $cm^{-1}$)[4,5], reasonably high charge-carrier mobility[6] and wide range of band-gap tunability (from 1.2 to 2.8 eV)[7,8]. Typically, HP-based photodetectors are thin-film devices with a microcrystalline absorber, sandwiched between charge-transporting layers of p-type and n-type materials. The use of p-i-n oriented HP based PDs allows to reach competitive output performance with detectivities in the range $10^{12}$-$10^{13}$ Jones[9,10] which a comparable with benchmarks in the industry of sensors like amorphous Si. The utilization of various solution-processing techniques (slot-die[11], ink-jet printing[12]) offers opportunities for low-cost industrial fabrication[13]. HP devices were considered 'defect-tolerant' materials[14] due to their reduced dynamics of non-radiative recombination processes[15]. However, various reports estimating the numerical parameters of trap states[16–18] highlight the presence of deep states with concentrations up to $10^{14}$ $cm^{-3}$. Structural imperfections, point defects, and impurities, typically associated with grain boundaries and interfaces in HP devices, can significantly affect device performance. The migration of ionic defects can lead to several detrimental effects, including hysteresis[19], charge accumulation[20], increased leakage current[21], longer rise times, and signal decay. Additionally, clusters of charged defects or decomposition products of the perovskite molecule can initiate electrochemical corrosion at interfaces[22], associated with oxidation processes. Researchers have attempted to mitigate these challenges with various materials for inter-grain passivation[23]. The parasitic trapping processes in HP photodetectors could occur in the perovskite (bulk recombination) and at the charge collection junction with p- and n-type transporting layers (surface recombination), respectively [24,25]. Most efforts to passivate surfaces in HP photodiodes (PPDs) focus on modifying a single interface[26–28].

The study conducted by *Ollearo et al.* revealed that modifying the hole transport interface with polymer thin-films yielded improvements in energy level alignment, leading to a decrease in dark current to $10^{-11}$ A/cm2 and noise equivalent power to $2 \times 10^{-14}$ A $Hz^{-1/2}$. *Wonsum Kim* and *colleagues*[29] passivated grain boundaries by adding a PMMA interlayer between the perovskite and ETL. This method has decreased the noise equivalent power (NEP) from 2520 $fW/Hz^{1/2}$ to 230 $fW/Hz^{1/2}$, but the response speed and cutoff bandwidth were also reduced. Improving device performance requires a more comprehensive approach. Forming heterostructures with wide-gap semiconductors and dielectrics is an effective strategy to enhance the performance of photodetectors (PDs). The presence of ultrathin dielectric layers at the interface can facilitate better charge extraction by modifying the electric field distribution within the photodiode[30]. Generally, the key advantages include improved carrier confinement, reduced band-bending, decreased interface trap-states density, and high-frequency capabilities. Notably, the effectiveness of semiconductor-ultrathin dielectric interfaces has been demonstrated in various configurations: high-K materials for GaAs devices (Hf-based)[31,32], conventional dielectrics ($Al_2O_3$, $Si_3N_4$)[32], and insulating polymers (polyimide)[33]. Perovskite photodiodes face a greater challenge in addressing this issue. It involves passivating the intergranular boundaries of the microcrystalline absorber and surface with selective transport layers that are prone to iodine-containing defects diffusion[34]. Among various polymer dielectric materials, which can be used for the passivation of perovskite-based devices the most interesting those based on polyvinylidene fluoride (PVDF), due to its unique ferroelectric and dielectric properties [35]. In general, ferroelectric materials can be polarized and generate a built-in electric field to separate the photogenerated electron-hole pairs. However, due to the high crystallinity and insufficient solubility of PVDF, its copolymers, especially those with trifluoroethylene, have become more widely used because their thin and uniform films can be easily obtained by standard solution methods[36–38]. Several groups have already described integration of poly(vinylidenefluoride-trifluoroethylene) (P(VDF-TrFE) into the perovskite absorber layer to manipulate the carrier transfer behavior of the devices[39,40]. Typically, research on the integration of dielectrics into photodiode structures considers only one type of application, either modifying the inter-grain interfaces in the absorber or at the contact with charge-transport layers. However, integrated interface modification using single P(VDF-TrFE) films for compensation of charged species in both cases may provide a synergistic effect. On one hand, this impacts the generation and recombination processes, while on the other hand, it affects the extraction and transport of photocarriers. The combinational impact of the dielectrics application in perovskite photodiodes remains sufficiently unexplored and requires deep investigation.

In this paper, we present a comprehensive study on the modification of the perovskite absorber and the n-type collection interface using a promising organic polymer dielectric/ferroelectric P(VDF-TrFE). In this work, we reveal the effects of P(VDF-TrFE) integration on the optoelectronic and surface properties of multicationic perovskite thin films and investigate the reasons for changes in charge carrier transport in p-i-n diode structures. For the first time, we demonstrate that the complex implementation of ultrathin dielectric layers improves the wide spectra of the photo-response parameters (specific detectivity, noise equivalent power, linear dynamic range), but also increases the fast response and cut-off frequency without special pre-conditioning. The obtained results are thoroughly analyzed and discussed.

**Results and discussions**

In this research, we developed a novel modification of the microcrystalline perovskite absorber ($Cs_{0.2}FA_{0.8}PbI_{2.93}Cl_{0.07}$) interfaces by utilizing P(VDF – TrFe). The copolymers of this material possess a combination of low dielectric constants with relatively high glass transition temperature, transparency, structural flexibility, ease of solution processing, and chemical stability [41]. We employed ultrathin P(VDF-

TrFE) dielectric interlayers at the grain boundaries within the perovskite film and at the absorber/electron transport layer interface. The schematics of HP based photodiodes (**PDs**) developed in this work and molecular structure of P(VDF-TrFE) presented in the **fig.1**.

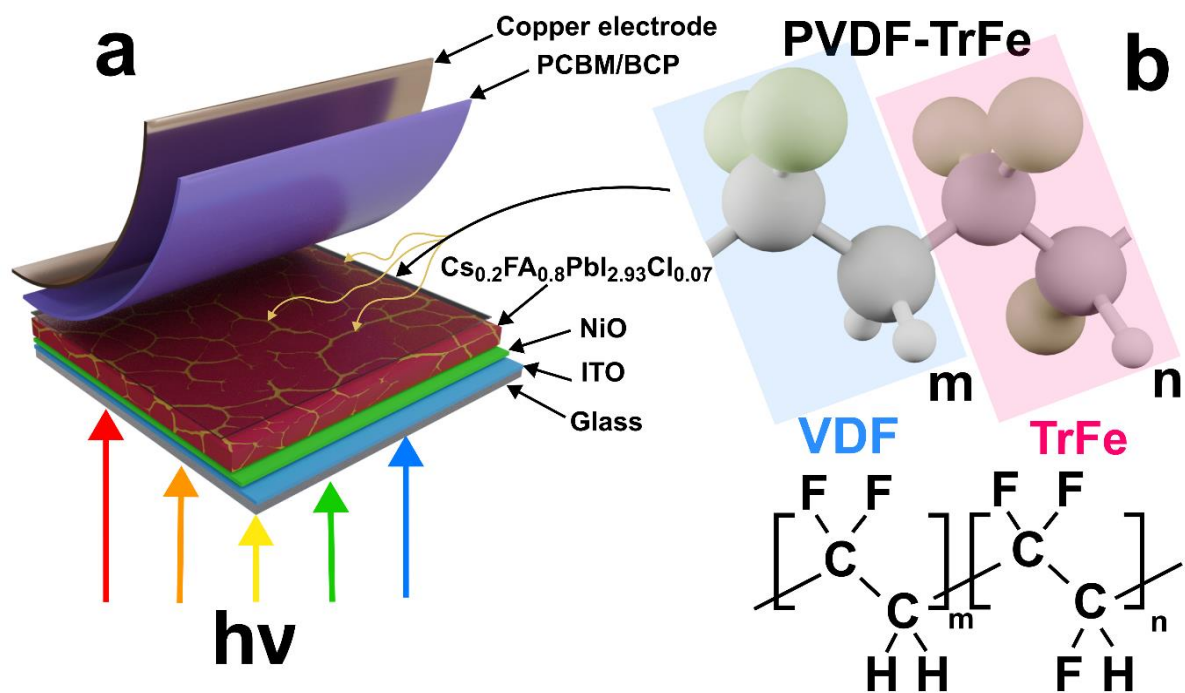

Figure 1 – Photodiode scheme (a) with P(VDF-TrFE) structure visualization and formula (b)

The devices were fabricated with a p-i-n architecture: glass (1.1 mm)/ITO (anode, 330 nm)/NiO (p-type, 30 nm)/perovskite absorber $Cs_{0.2}FA_{0.8}PbI_{2.93}Cl_{0.07}$ (450 nm)/PCBM (n-type, 30 nm)/BCP (hole blocking interlayer, 10 nm)/Copper (cathode, 100 nm). The solubility of P(VDF-TrFE) in solvents like dimethylformamide and N-methyl pyrrolidone allowed its integration into the perovskite solution for subsequent deposition and crystallization. Through this technological approach, we achieved a bulk distribution of the dielectric within the microcrystalline absorber. P(VDF-TrFE) deposition on the back surface of $Cs_{0.2}FA_{0.8}PbI_{2.93}Cl_{0.07}$, was also carried out by solution-processing. A detailed description of the experimental section presented in the Electronic Supplement Information file (**ESI**). Fabrication of the different types of samples was associated with modifying the perovskite absorber interfaces. To simplify the titles of the different configurations, we designated the bare perovskite films as "control". The sample with an P(VDF-TrFE) distributed into the bulk of $Cs_{0.2}FA_{0.8}PbI_{2.93}Cl_{0.07}$ was designated as "Bulk", and the sample with dielectric incorporated to the bulk and electron collection junction was referred as "Bulk/n-side".

We examined the optical properties of perovskite layers with dielectric interlayers using Tauc plots and photoluminescence spectra (**fig.2**). The calculation was realized using equations **S1** (**ESI**). The analysis of absorption properties analyses for thin-films demonstrated edge shifts (**fig.2(a)**). The band-gap energy ($E_g$) was extracted using linear approximation, revealing that the $E_g$ of the control sample was 1.573 eV. Integrating dielectric interlayers led to a decrease of the $E_g$. The band-gap value for Bulk sample was 1.571 eV, while for Bulk/n-side the $E_g$ reduced to 1.570 eV. The analysis of peak positions in the photoluminescence spectra (**fig.2(b)**) correlated with the absorption data. Thin films containing dielectric interlayers exhibited a 'red shift.' The peak positions were 1.592 eV (779 nm) for the control, 1.583 eV (783 nm) for Bulk, and 1.581 eV (784 nm) for Bulk/n-side.

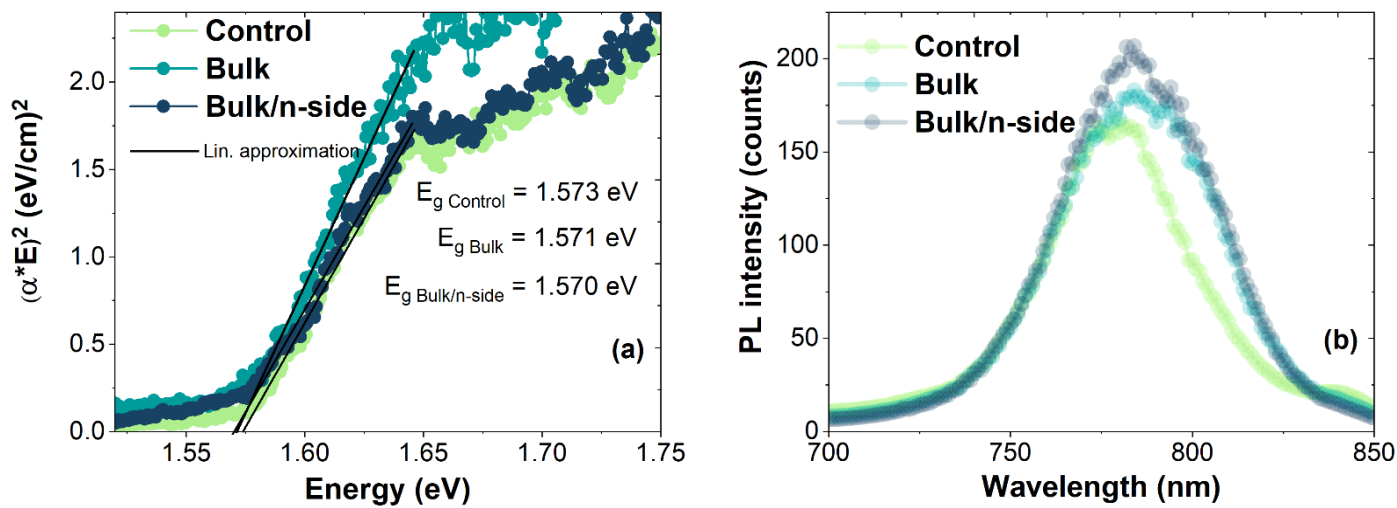

Figure 2 – Absorption spectra (a) and PL spectra (b) of obtained perovskites

The influence of P(VDF-TrFE) on perovskite films' morphology and phase composition was investigated through AFM (**Fig.3(a-c)**) and XRD (**Fig.3d**). Analysis revealed that bare perovskite absorber and P(VDF-TrFE) containing samples predominantly exhibited the β-$Cs_{0.2}FA_{0.8}PbI_{2.93}Cl_{0.07}$ phase, characterized by specific peaks at 20.6º, 29.3º, 36.14º, 41.9º, and 47.15º. Furthermore, distinct peaks corresponding to the ITO substrate were observed, along with peaks signifying lead iodide at 18.5° (001) and 39.2° (101). No signs of peak shifting or broadening were detected. So, integration of the P(VDF-TrFE) hadn't discernible quantitative or qualitative effect on the phase composition of the perovskite layer. To examine the impact of polymers on the optoelectronic properties of perovskite, KPFM measurements were conducted using an NT-MDT Ntegra AFM equipped with a NSG03/Au tip. The work function ($W_f$, **Fig. 3e**) was determined with a fresh HOPG surface serving as the energy baseline (see details of the calculations in the **eq.S2,S3** in **ESI**). The Fermi energy of the control sample is 4.85 eV. Incorporating PVDF into the perovskite volume shifts the Fermi level to 4.16 eV. Applying PVDF to the bulk sample increased the work function from 4.16 eV to 4.28 eV. Both bulk and bulk/n-side samples displayed a significant decrease in work function, up to 0.7 eV for the bulk sample, compared to the control perovskite film.

The incorporation of P(VDF-TrFE) into the bulk and at the interfaces of perovskite can induce the reconfiguration of the trap states. P(VDF-TrFE), a ferroelectric polymer, has been shown to influence the electronic properties of perovskite materials by reconfiguring intrinsic defects and trap states, leading to improved charge separation and reduced recombination. The shift in work function from 4.85 eV (intrinsic conditions) to 4.28 eV for the Bulk/n-side sample indicates an alignment favorable for electron transport, which could be attributed to the modification of trap states at the interface and within the bulk of the perovskite. The observed Fermi level pinning (FLP)[42–44] effect may be caused by the ferroelectric properties of P(VDF-TrFE). The enhancement of local fields and the formation of dipoles at intergranular boundaries can screen or compensate for states induced by ionic defects, whose concentration in $Cs_{0.2}FA_{0.8}PbI_{2.93}Cl_{0.07}$ thin-films can reach up to $10^{14}$ cm$^{-3}$[45].

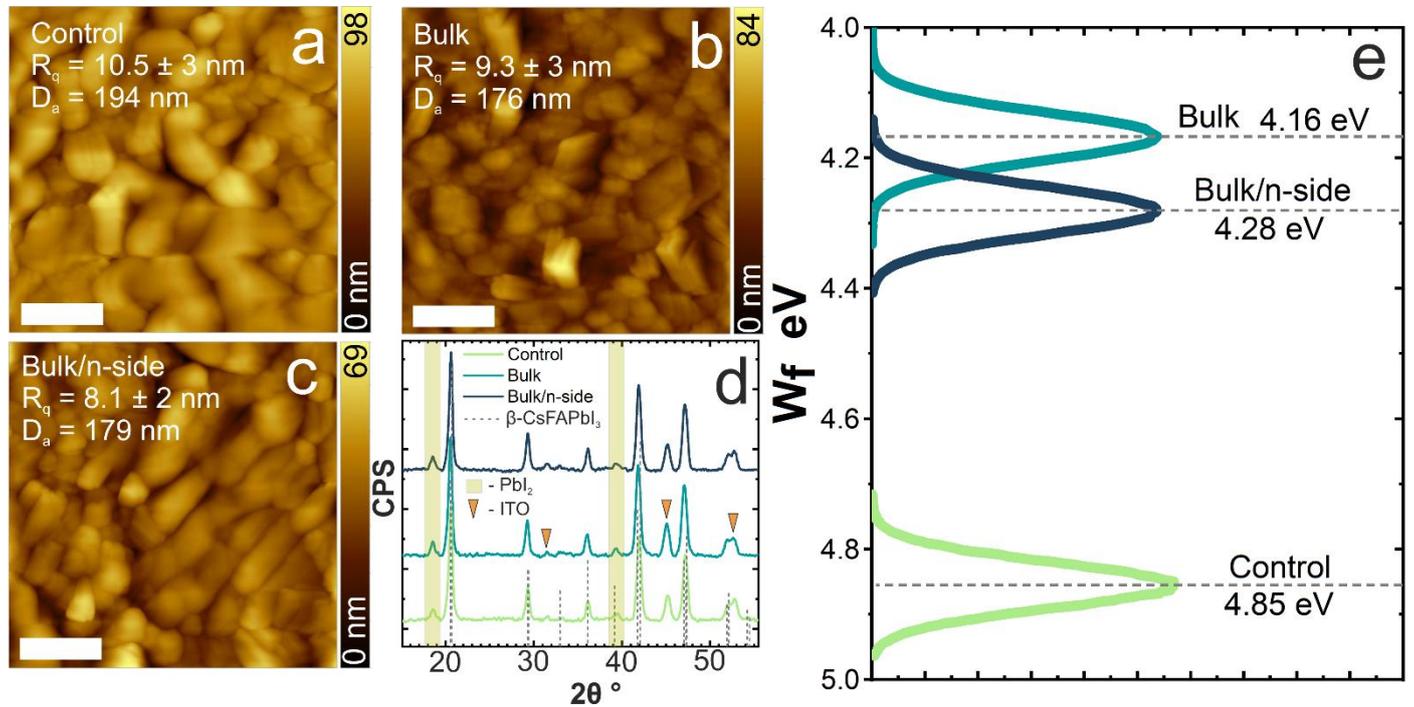

Figure 3 – morphology (a-c), X-Ray diffraction (d) and surface work function of perovskite films (e); sale bar 500 nm

To evaluate the recombination dynamics in thin films with dielectric interlayers, we used high time-resolved luminescence (TRPL) measurements (**fig.S1** in **ESI** and **tab.1**). Photo-carrier lifetime values were extracted by fitting eq.: $I(t)=A_1\exp(-x/\tau_1)+A_2\exp(-x/\tau_2)$ (double exponent fit). For relevant data, we collected batch statistics from at least six samples for the control, bulk, and bulk/n-side samples. The median lifetime ($\tau$) of the control CsFAPbI$_3$ was 72.6 ns. The integration of P(VDF-TrFE) into the intergranular boundaries minimally changed $\tau$ to 74.6 ns. For the Bulk/n-side configuration, a more substantial increase in $\tau$ to 82.6 ns (+14% compared to the control) was observed. Interpreting TRPL data is a complex task. For the fast component, the $\tau_1$ value decreased from 8.9 ns to 6.9 ns with P(VDF-TrFE) integration, potentially indicating a reduced contribution of trap-assisted recombination [46,47]. For the slow component, $\tau_2$, which may be influenced by exciton states[48,49], we observed inverse correlations. We assume that this trend could result from the local polarization of P(VDF-TrFe) during the photo-injection of electron-hole pairs and the resulting accumulation effects in the enhanced electric field.

**Table 1 -** TRPL normalized data for perovskite films NiO/Control; NiO/Bulk; NiO/Bulk/n-side

| Sample configuration | $A_1$, $\div 10^3$ | $\tau_1$ (ns) | $A_2$, $\div 10^3$ | $\tau_2$ (ns) | $\tau_{average}$ (ns) |
| --- | --- | --- | --- | --- | --- |

| | | | | | | |
|---|---|---|---|---|---|---|
| **Control** | median | 6.9 | 8.9 | 20.7 | 75.1 | 72.6 |
| | average | 7.2 | 8.6 | 21.0 | 77.4 | 74.9 |
| | ± std | ± 0.8 | ± 1.1 | ± 4.3 | ± 14.8 | ± 14.7 |
| **Bulk** | median | 6.3 | 8.1 | 21.6 | 76.6 | 74.6 |
| | average | 6.3 | 8.2 | 21.2 | 76.6 | 74.5 |
| | ± std | ± 0.6 | ± 0.8 | ± 1.4 | ± 2.4 | ± 2.1 |
| **Bulk/n-side** | median | 7.9 | 6.9 | 20.3 | 81.6 | 79.2 |
| | average | 7.8 | 6.9 | 20.5 | 85.0 | 82.6 |
| | ± std | ± 0.7 | ± 0.4 | ± 1.3 | ± 8.2 | ± 8.0 |

*Notes: $A_1$ and $A_2$ is intensity amplitudes; $\tau_1$ and $\tau_2$ is luminescence lifetimes; $\tau_{average}$ is a weighted average luminescence lifetime; std is standard deviation.*

To investigate the impact of the interface modification on diode properties, we measured the dark volt-ampere characteristics (JVs) for the fabricated PPDs with bias range from -0.1 V to +1.1 V, as shown in **Fig.4**. The dark JV curves of the fabricated photodiodes exhibited typical diode behavior, characterized by relevant rectification. Generally, the dark JV curve can be divided into four main regions related to the shunt current (I), recombination current (II), diffusion current (III), and contact resistance (IV). We observed that integrating the dielectric both within the bulk and at the n-type interface altered the diode properties, indicating notable changes in the charge carrier transport. The minimum value of the dark current density ($J_{min}$) measured at zero bias was $1\times10^{-7}$ A/cm² for the reference device. In contrast, the values for the Bulk and Bulk/n-side configurations were significantly reduced by an order of magnitude, up to $2\times10^{-8}$ A/cm² and $7\times10^{-10}$ A/cm², respectively. The use of P(VDF-TrFE) further contributed to reducing the dark leakage current ($J_{leakage}$, bias= -0.1 V). The control and bulk PPDs exhibited comparable $J_{leakage}$ values ~$2\times10^{-6}$ A/cm², whereas the bulk/n-side configuration demonstrated a significant reduction to $9.7\times10^{-8}$ A/cm². Calculating the shunt resistance ($R_{sh}$) values showed an increase for devices with integrated dielectrics at the interfaces. The reference device had an $R_{sh}$=4.2 kOhm·cm², which increased to 6.8 kOhm·cm² for the bulk configuration and reached 1.2 MOhm·cm² for the Bulk/n-side device. For a comprehensive analysis of the changes in the dark JV curve in the II and III regions, we performed a fitting using a 2-diode model to extract the appropriate parameters (**tab.2** with values of the non-ideality factors ($m_1$, $m_2$); reverse saturation currents ($J_{01}$, $J_{02}$), and shunt resistance).

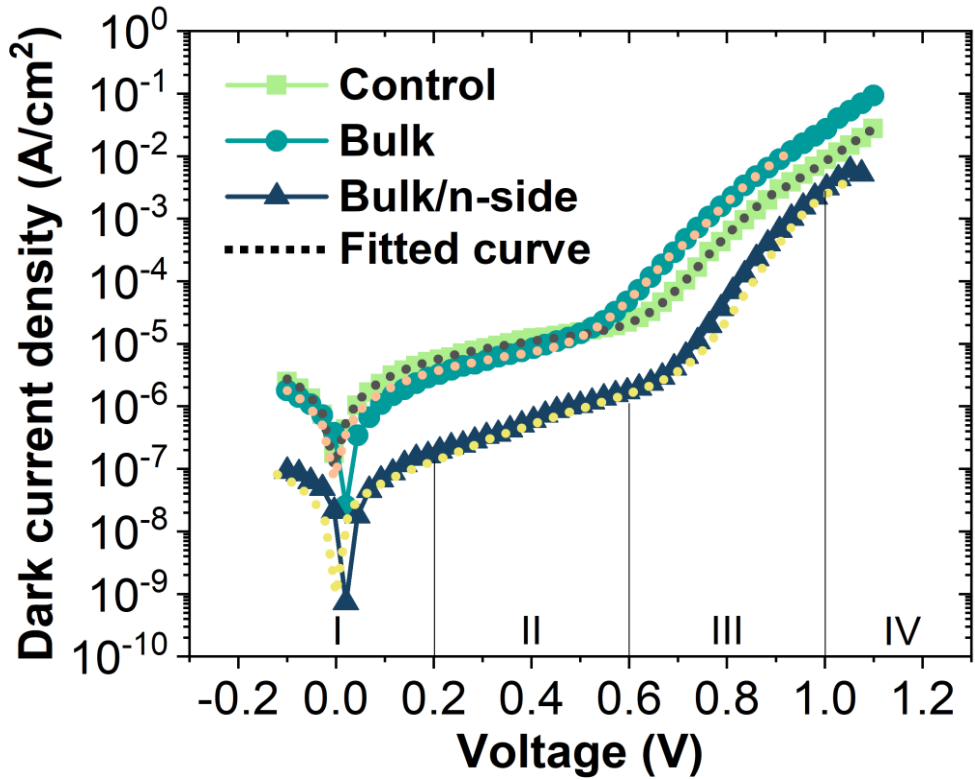

Figure 4 – Dark JV curves of p-i-n structures

Generally, the total saturation dark current ($J_0$) is governed by recombination processes within the solar cell and can be utilized to estimate the efficiency of charge carrier transport. In hetero-structured devices, $J_0$ depends on the number and quality of interfaces. In our study, we modified the interfaces within the perovskite absorber volume and at the junction for electron collection. Calculations for dark JV were performed according to the equations **S4-S11 (ESI)**, the equivalent circuit of the 2-diode model presented in **ESI (fig.S2)**. The calculated values of $J_0$ indicated that the overall value is primarily determined by $J_{02}$. The significant reduction of $J_0$ from $10^{-4}$ to $10^{-8}$ A/cm² in devices with P(VDF-TrFE) integration compared to control highlights the reduced contribution of recombination to charge carrier transport. The non-ideality factor of a photodiode is typically used to analyze the dominant recombination mechanisms involving free carriers and traps both in the volume (bulk) and at interfaces. Interpreting the calculated values of the non-ideality factor for photovoltaic semiconductor devices is a complex task and varies greatly depending on the photodiode architecture. In a classical p-n junction, the non-ideality factor (**m**) equals 1 when diffusion current predominates over recombination current. The parameter can reach a value of 2 when recombination current dominates in the space charge region. A lower non-ideality factor indicates reduced recombination dynamics and increased efficiency of the junction for charge carrier transport. A lower non-ideality factor points to reduced recombination dynamics and increased junction efficiency for charge carrier transport. P-i-n diode devices based on halide perovskites are more complex, involving two hetero-junctions. Therefore, recombination processes in PPDs are not described by the *m* range of 1-2 and require modification of standard equivalent electrical circuits. In our work, we used a 2-diode model with diodes in series, calculating the non-ideality factor as the sum of the values of each diode. For the control sample, the calculated non-ideality factor was 2.811. In the Bulk configuration, the value significantly decreased to 1.792, demonstrating notable changes in recombination processes associated with the interfaces of the microcrystalline perovskite film. Interestingly, the non-ideality factor for the Bulk/n-type configuration, while lower than the control at 2.610, exceeded the Bulk device values. Potential losses in charge carrier transport efficiency could be attributed to the non-optimized P(VDF-TrFE) deposition process on the surface.

Table 2. Calculated parameters for fitted dark JV curves

| Device configuration | $m_1$ | $m_2$ | $m$ ($m_1+m_2$) | $J_{01}$, A*cm$^{-2}$ | $J_{02}$, A*cm$^{-2}$ | $R_{sh}$, ohm*cm$^2$ |
|---|---|---|---|---|---|---|
| Control | 1.534 | 1.277 | 2.811 | 1.34E-12 | 8.78E-4 | 3.66E+04 |
| Bulk | 1.112 | 0.680 | 1.792 | 1.72E-13 | 3.17E-6 | 5.58E+04 |
| Bulk/n-type | 1.320 | 1.290 | 2.610 | 3.03E-11 | 6.66E-8 | 1.62E+06 |

The photo-response of PPDs in short-circuit and open-circuit modes was assessed over a wide range of illumination power densities ($P_0$, 540 nm, LED source, range: $10^{-3}$ to $10^1$ mW/cm²). For the short-circuit current ($J_{sc}$), linear trends were observed, as the photocurrent $\propto$ charge generation rate. The calculation of the responsivity (**R, eq.S12** in **ESI**), representing photocurrent dependence on $P_0$ at a wavelength of 540 nm, resulted in values of 0.32 A/W for the control device, 0.38 A/W for the Bulk, and 0.44 A/W for the Bulk/n-side configuration. The dependence of open-circuit voltage ($V_{oc}$) vs. $P_0$ exhibited a logarithmic nature. The Bulk/n-side modification of the device increased $V_{oc}$ across the full range of light intensities by 0.02 to 0.05 V compared to control PPDs. On the other hand, Bulk devices showed the highest $V_{oc}$ values at low-light intensities (<$10^{-1}$ mW/cm²), indicating changes in the separation and collection of the charge carriers.

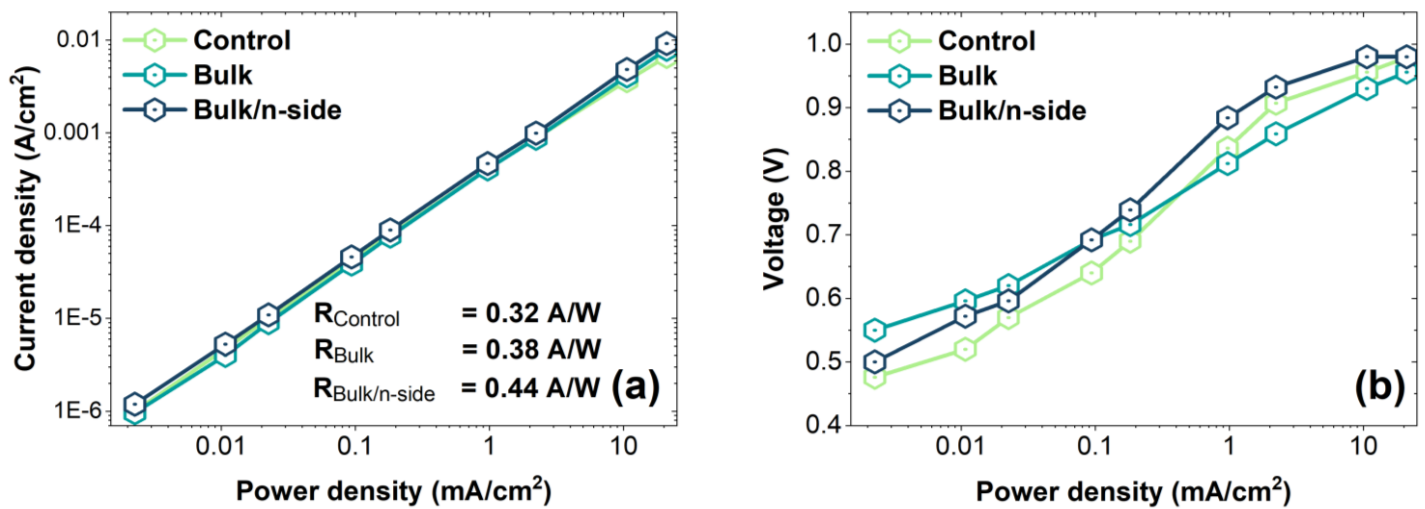

Figure 5 – The linearity plot of $J_{sc}$ to $P_0$ for the various perovskites modifications on a logarithmic scale (a); the dependence of $V_{oc}$ vs. $P_0$ on a semi-logarithmic scale (b)

For the illumination power density range used, we calculated the linear dynamic range (LDR, eq. **S13** in ESI), which defines the range of incident light power over which the photodiode output current depends linearly on the input optical power. This critical characteristic determines the photodiode's ability to accurately measure light intensity over a wide range of power levels. Calculations of LDR values under 540 nm LED illumination indicated an increase from 76.6 dB to 99.5 dB for PPDs with P(VDF-TrFE) interlayers (**tab.3**). This originated from the reduction in dark current and the improvement in the signal-to-noise ratio for the Bulk and Bulk/n-side configurations. Across the $P_0$ range from $10^{-3}$ to $10^1$ mW/cm², we observed no deviation from linearity in the photocurrent, suggesting that the LDR values for the designed PPDs may potentially exceed 100 dB. The spectral performance of the devices was estimated via measurements of the external quantum efficiency (**fig.S3** in **ESI**). All PPD configurations exhibited a high level of photoelectric conversion, between 82% and 87%, in the visible part of the spectrum. This correlates with the data obtained under high optical pumping conditions (540 nm).

Based on the assumption that shot noise is dominant in PPDs, we estimated the key figures of merit parameters—specific detectivity (**D\***) and noise equivalent power (**NEP**)—for the short-circuit mode (zero bias). The data calculated using **eq. S14, S15** (**ESI**) are presented in **tab.3**. Reducing current leakage in devices with modified interfaces increased D* by almost an order of magnitude, approaching the value of $10^{12}$ Jones for the Bulk/n-side configuration. Strengthening the shunt properties and increasing R led to a reduction in the NEP for the P(VDF-TrFE)-containing devices. While the NEP for the control sample was around $\sim 10^{-12}$ W·Hz$^{-1/2}$, we obtained $6.6 \times 10^{-13}$ and $4.00 \times 10^{-13}$ W·Hz$^{-1/2}$ for the Bulk and Bulk/n-side configurations, respectively.

Table 3. Calculated D* and NEP for the fabricated PPDs

| Device | LDR (dB) | D* (Jones) at 540 nm | NEP (W·Hz$^{-1/2}$) |
| --- | --- | --- | --- |
| Control | 76.6 | $2.20 \cdot 10^{11}$ | $1.76 \cdot 10^{-12}$ |
| Bulk | 92.0 | $5.78 \cdot 10^{11}$ | $6.65 \cdot 10^{-13}$ |
| Bulk/n-side | 99.5 | $9.69 \cdot 10^{11}$ | $4.00 \cdot 10^{-13}$ |

To assess the response dynamics of PPDs, we measured the transient characteristics of ON/OFF modes during optical pumping (**fig.6**). A square pulse of light (LED, 50 kHz) was used to evaluate changes in the rise and fall profiles of the photocurrent over time. Specifically, we analyzed the rise and fall times ($t_r$ and $t_f$) of the current response at amplitudes corresponding to 10% and 90% of signal saturation. The devices exhibited a characteristic π-shaped waveform. For the control device, the $t_r$ was 6.3 μs, and the $t_f$ was 10.9 μs. Bulk PPDs demonstrated enhanced performance with decreased transient times: $t_r$ = 4.9 μs (-22.6%) and $t_f$ = 6.7 μs (-15.6%). The Bulk/n-side configurations showed the most rapid signal growth profile with $t_r$ = 4.6 μs (-32.6%), while decay was 6.5 μs.

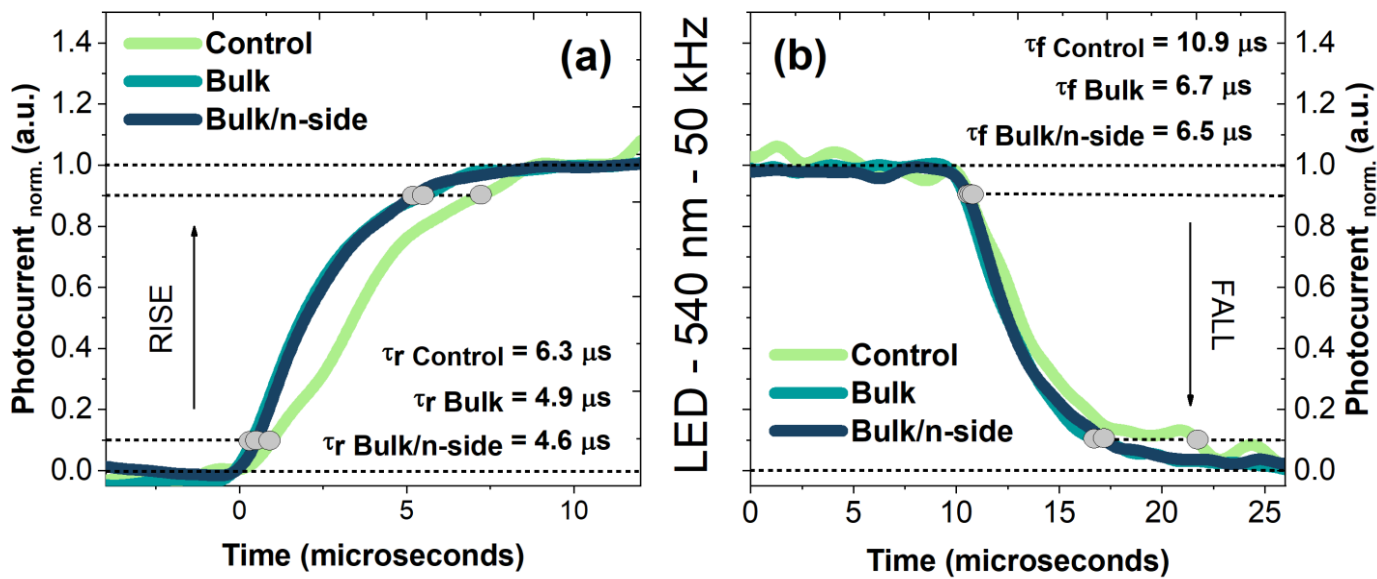

Figure 6 – Response speed of the control device (a); device with PVDF in perovskite volume (b) and device with PVDF both in perovskite volume and on the interface with n-transport layer (c)

A crucial parameter of the photodiode related to photo-response dynamics is the cut-off frequency, defined as the frequency at which the photodiode's response diminishes to 3 dB below its maximum value (**f$_{3dB}$**). This parameter determines the speed at which the photodiode can respond to the relevant changes in the light signal. The 3 dB point corresponds to a power reduction of half of its maximum value or a decline in current to 0.707 from the maximum. Essentially, f$_{3dB}$ sets the upper limit on the frequency of the light signal that the photodiode can efficiently detect. For f$_{3dB}$ measurements, the amplitude of the photodiode photocurrent was measured when optically pumped with a square pulse of a green LED (540 nm) at different frequencies of the LED glow. Measurements were carried out for the frequency range from 10 to 90 kHz. The illumination was carried at a power density of 2 mW/cm$^2$. The measured data are presented in **fig.7**. Enhanced speed and transport performance in the PPDs, containing P(VDF-TrFE), increased the f$_{3db}$ from 64.8 kHz for the control device to 71.9 kHz and 76.8 kHz for the Bulk and Bulk/n-side configurations, respectively. To evaluate the stability of the devices, we measured the transient characteristics after 250 hours of thermal stress (70°C) (**fig.S4** in **ESI**). For all PPDs configurations decrease in response speed was negligible.

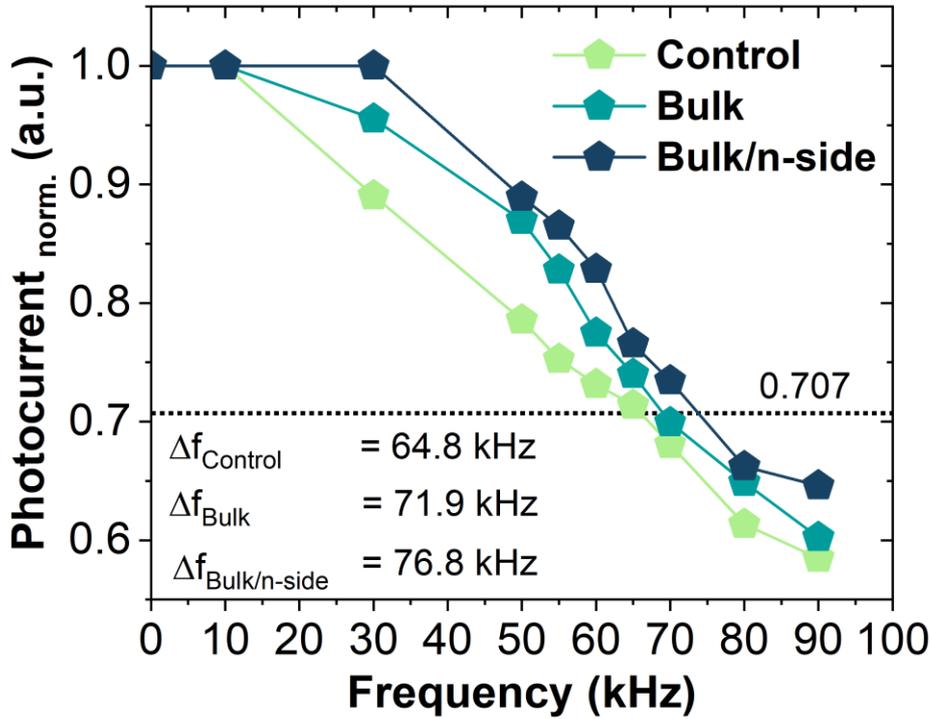

Figure 7 – $F_{3db}$ bandwidth for the fabricated PPDs with dielectric interlayers

The integration of P(VDF-TrFE) dielectric materials has been demonstrated to be effective for perovskite-based diode structures. In early works [50,51], researchers utilized P(VDF-TrFE) to enhance the built-in electric field within perovskite absorbers, thereby improving the collection efficiency of the free charge carriers. However, the previously presented approaches relied on poling effects, which potentially initiate ion migration processes under polarization conditions. However, poling and migration effects can adversely affect device performance. In our work, we have demonstrated a novel approach involving complex modification of interfaces within the bulk and at the electron collection junction, wherein P(VDF-TrFE) primarily compensates for charge carrier traps. PPDs-containing incorporating P(VDF-TrFE) in the bulk absorber and at bulk/n-type interfaces operated in regimes similar to reference devices and didn't require pre-conditioning. We observed a relevant increase in photo-carrier lifetime with the integration of P(VDF-TrFE) on the perovskite film surface, indicating a influence of the dielectric interlayer on recombination processes and charge carrier splitting. A comprehensive analysis of diode transport characteristics also shows a reduced contribution of non-radiative recombination at the interfaces for the devices with dielectric interlayers. P(VDF-TrFE) plays a significant role in enhancing the shunt properties of PPDs and reducing dark currents. The presence of a thin dielectric layer assists in mitigating potential defects in morphology, such as micro-pinholes. This correlates with the measured photo-response of PPDs. The increase in **R** is linked to better photo-carrier collection efficiency (see **eq.S16** in **ESI**)[52]. The rise in $V_{oc}$ in low light intensity regions results from improved energy level alignment and reduced trap-states in the band-gap, which negatively impact quasi-Fermi level splitting[53] during photo-injection. Improved dynamic performance is a key result of this work. The photo-response profile directly correlates with the cut-off frequency. Different capacitive effects according to multiple models [19,54,55] describe variations in transition time in perovskite diode architectures. The appearance of slow components in the transient photocurrent (as observed in the PPD control) is typically associated with the accumulation of ionic species at interfaces [56–58], which in a perovskite photodiode can be represented by iodine vacancies, uncompensated organic cation ions, iodine in interstitials, etc. The photodiode junction capacitance, combined with other capacitances in the circuit, forms an RC constant with the load resistance, which limits the device's bandwidth. The charge-carrier transit time and RC constant determine the cut-off frequency (**eq.1**) [59–61]:

$$f_{-3dB}^{-2} = \left(\frac{3.5}{2\pi t_{CCT}}\right)^{-2} + \left(\frac{1}{2\pi RC}\right)^{-2} \tag{1}$$

Where $t_{CCT}$ - charge-carrier transit time;

R - total series resistance, including the device resistance, contact resistances, and load resistances;

C - the sum of the capacitance of the device.

The improvements in dynamic response for the Bulk/n-side configuration result from a complex interplay of suppressed trapping, reduced accumulation effects, and minimized current leakage. Compared to similar studies on integrating dielectrics into the absorber volume and interfaces of a perovskite photodiode[51,62–64], our results demonstrate superior f₃dB and improved transition times.

**Conclusions**

In this work, we performed a comprehensive study of P(VDF-TrFE) integration in bulk of the absorber and electron transport surfaces for PPDs. We demonstrated that dielectric incorporation provides Fermi level pinning, shifting it from 4.85 eV for bare $CsFAPbI_3$ to 4.16-4.28 eV. This adjustment reconfigures the energy level alignment that potentially allows more efficient electron collection. The Bulk/n-side configuration exhibited beneficial improvements in the entire spectra of photodiode performance parameters. This underscores the benefits of a complex modification of both the microcrystalline absorber and hetero boundaries with standard transport layers like $C_{60}$ fullerenes. The dielectric interface layers improved diode properties by reducing the non-ideality factor and dark leakage currents. The responsivity of the fabricated PPDs was 0.32 A/W for the control device, 0.38 A/W for the Bulk, and 0.44 A/W for the Bulk/n-side configuration. Reducing the impact of non-radiative recombination processes at the interfaces allowed to decrease the NEP to ~$10^{-13}$ W·$Hz^{-1/2}$ and achieve a relevant D* of ~$10^{12}$ Jones. Despite the meaningful differences in the photo response ($I_{sc}$ vs $P_o$, $V_{oc}$ vs $P_0$) between bulk and Bulk/n-side devices, the dynamic response showed close behavior. The cut-off frequency was increased from 64.8 (control) to 71.9 (Bulk) and 76.8 kHz (Bulk/n-side). Photodiodes with f3dB values between 50 - 100 kHz are considered for use in applications that don't require extra fast response (GHz range), such as light measurement and sensing in industrial or scientific equipment, optical switches and encoders, low-speed optical communication systems, some types of the optical remote controls. This paper demonstrates the importance of complex interface processing in perovskite photodiodes, with a focus on the unique bulk modification and heterojunction with electron-transport layers. The developed method of using dielectrics combines the efficient tuning of transport properties and fast performance. Our work provides new insights into interface engineering for perovskite optoelectronics, aiming to comprehensively enhance the output performance.

**SUPPLEMENTARY MATERIAL**

The supplementary material includes description of the materials section (materials, inks preparation, device fabrication, characterization), details for the double diode model used for fitting of dark JV curves, external quantum efficiency spectra, and transient response after heat-stress and ON/OFF cycles.

The data that supports the findings of this study are available within the article and its supplementary material

**Acknowledgments**

The authors gratefully acknowledge the financial support from Russian Science Foundation with project № 22-19-00812.

**Author contributions**

**D.S.S. and Yu.N.L.** conceived the work.

**Yu.N.L. and N.S.S.** provided organic dielectric materials.

**A.P.M., S.Yu.U., P.A.G., S.I.D.** performed device characterization and IV modelling

**L.O.L., D.S.M., N.S.S.** performed advanced characterization of surface and optical properties.

**D.S.S. and Yu.N.L.** provided administrative support and resources.

The manuscript was written with contributions from all the authors. All the authors approved the final version of the manuscript.

### References

[1] Y. Wu, J. Feng, Z. Yang, Y. Liu, S. (Frank) Liu, *Advanced Science* **2023**, *10*, DOI 10.1002/advs.202205536.

[2] H. Wang, Y. Sun, J. Chen, F. Wang, R. Han, C. Zhang, J. Kong, L. Li, J. Yang, *Nanomaterials* **2022**, *12*, 4390.

[3] N. Yaghoobi Nia, D. Saranin, A. L. Palma, A. Di Carlo, in *Solar Cells and Light Management*, Elsevier, **2020**, pp. 163–228.

[4] Z.-K. Tang, Z.-F. Xu, D.-Y. Zhang, S.-X. Hu, W.-M. Lau, L.-M. Liu, *Sci Rep* **2017**, *7*, 7843.

[5]   Z. Chen, Q. Dong, Y. Liu, C. Bao, X. Xiao, Y. Bai, Y. Deng, J. Huang, Y. Fang, Y. Lin, S. Tang, Q. Wang, *Nat Commun* **n.d.**, 1.

[6]   C. Motta, F. El-Mellouhi, S. Sanvito, *Sci Rep* **2015**, DOI 10.1038/srep12746.

[7]   S. A. Kulkarni, T. Baikie, P. P. Boix, N. Yantara, N. Mathews, S. Mhaisalkar, *J. Mater. Chem. A* **2014**, *2*, 9221.

[8]   A. Amat, E. Mosconi, E. Ronca, C. Quarti, P. Umari, Md. K. Nazeeruddin, M. Grätzel, F. De Angelis, *Nano Lett* **2014**, *14*, 3608.

[9]   R. Ollearo, A. Caiazzo, J. Li, M. Fattori, A. J. J. M. van Breemen, M. M. Wienk, G. H. Gelinck, R. A. J. Janssen, *Advanced Materials* **2022**, *34*, DOI 10.1002/adma.202205261.

[10]  D. Wu, W. Li, H. Liu, X. Xiao, K. Shi, H. Tang, C. Shan, K. Wang, X. W. Sun, A. K. K. Kyaw, *Advanced Science* **2021**, *8*, DOI 10.1002/advs.202101729.

[11]  T. S. Le, D. Saranin, P. Gostishchev, I. Ermanova, T. Komaricheva, L. Luchnikov, D. Muratov, A. Uvarov, E. Vyacheslavova, I. Mukhin, S. Didenko, D. Kuznetsov, A. Di Carlo, *Solar RRL* **2022**, *6*, 2100807.

[12]  L. Zhang, S. Chen, X. Wang, D. Wang, Y. Li, Q. Ai, X. Sun, J. Chen, Y. Li, X. Jiang, S. Yang, B. Xu, *Solar RRL* **2021**, *5*, 2100106.

[13]  M. Cai, Y. Wu, H. Chen, X. Yang, Y. Qiang, L. Han, *Advanced Science* **2017**, DOI 10.1002/advs.201600269.

[14]  R. E. Brandt, J. R. Poindexter, P. Gorai, R. C. Kurchin, R. L. Z. Hoye, L. Nienhaus, M. W. B. Wilson, J. Alexander Polizzotti, R. Sereika, R. Žaltauskas, L. C. Lee, J. L. MacManus-Driscoll, M. Bawendi, V. Stevanović, T. Buonassisi, *Chemistry of Materials* **2017**, *29*, 4667.

[15]  P. Azarhoosh, S. McKechnie, J. M. Frost, A. Walsh, M. van Schilfgaarde, *APL Mater* **2016**, *4*, 091501.

[16]  K. Xue, C. Renaud, P. Y. Chen, S. H. Yang, T. P. Nguyen, in *Lecture Notes in Networks and Systems*, Springer, **2019**, pp. 204–209.

[17]  A. A. Vasilev, D. S. Saranin, P. A. Gostishchev, S. I. Didenko, A. Y. Polyakov, A. Di Carlo, *Optical Materials: X* **2022**, *16*, 100218.

[18]  A. S. Shikoh, A. Y. Polyakov, N. B. Smirnov, I. V. Shchemerov, D. S. Saranin, S. I. Didenko, D. V. Kuznetsov, A. Agresti, S. Pescetelli, A. Di Carlo, *ECS Journal of Solid State Science and Technology* **2020**, *9*, 065015.

[19]  C. Gonzales, A. Guerrero, J. Bisquert, *The Journal of Physical Chemistry C* **2022**, *126*, 13560.

[20] J. S. Park, J. Calbo, Y. K. Jung, L. D. Whalley, A. Walsh, *ACS Energy Lett* **2019**, DOI 10.1021/acsenergylett.9b00840.

[21] Z. Ni, H. Jiao, C. Fei, H. Gu, S. Xu, Z. Yu, G. Yang, Y. Deng, Q. Jiang, Y. Liu, Y. Yan, J. Huang, *Nat Energy* **2022**, *7*, 65.

[22] D. Di Girolamo, F. Matteocci, F. U. Kosasih, G. Chistiakova, W. Zuo, G. Divitini, L. Korte, C. Ducati, A. Di Carlo, D. Dini, A. Abate, *Adv Energy Mater* **2019**, *9*, 1901642.

[23] B. Chen, P. N. Rudd, S. Yang, Y. Yuan, J. Huang, *Chem Soc Rev* **2019**, DOI 10.1039/c8cs00853a.

[24] H. Uratani, K. Yamashita, *Journal of Physical Chemistry Letters* **2017**, *8*, 742.

[25] T. S. Sherkar, C. Momblona, L. Gil-Escrig, J. Ávila, M. Sessolo, H. J. Bolink, L. J. A. Koster, *ACS Energy Lett* **2017**, *2*, 1214.

[26] D. Wang, W. Xu, L. Min, W. Tian, L. Li, *Adv Mater Interfaces* **2022**, *9*, DOI 10.1002/admi.202101766.

[27] J. Lu, H. Wang, T. Fan, D. Ma, C. Wang, S. Wu, X. Li, *Nanomaterials* **2022**, *12*, DOI 10.3390/nano12122065.

[28] Y. Zhao, S. Jiao, S. Liu, Y. Jin, S. Yang, X. Wang, T. Liu, H. Jin, D. Wang, S. Gao, J. Wang, *J Alloys Compd* **2023**, *965*, DOI 10.1016/j.jallcom.2023.171434.

[29] W. Kim, J. Park, Y. Aggarwal, S. Sharma, E. H. Choi, B. Park, *Nanomaterials* **2023**, *13*, 619.

[30] M. Choi, G. Kang, D. Shin, N. Barange, C.-W. Lee, D.-H. Ko, K. Kim, *ACS Appl Mater Interfaces* **2016**, *8*, 12997.

[31] G. He, X. Chen, Z. Sun, *Surf Sci Rep* **2013**, *68*, 68.

[32] F. S. Aguirre-Tostado, M. Milojevic, K. J. Choi, H. C. Kim, C. L. Hinkle, E. M. Vogel, J. Kim, T. Yang, Y. Xuan, P. D. Ye, R. M. Wallace, *Appl Phys Lett* **2008**, *93*, DOI 10.1063/1.2961003.

[33] D. H. Lee, S. S. Li, S. Lee, R. V. Ramaswamy, *IEEE Trans Electron Devices* **1988**, *35*, 1695.

[34] A. Yakusheva, D. Saranin, D. Muratov, P. Gostishchev, H. Pazniak, A. Di Vito, T. S. Le, L. Luchnikov, A. Vasiliev, D. Podgorny, D. Kuznetsov, S. Didenko, A. Di Carlo, **2022**, *2201730*, 1.

[35] Prateek, V. K. Thakur, R. K. Gupta, *Chem Rev* **2016**, *116*, 4260.

[36] A. D. Khudyshkina, Yu. N. Luponosov, V. G. Shevchenko, S. A. Ponomarenko, *Express Polym Lett* **2021**, *15*, 957.

[37] E. A. Kleimyuk, A. I. Kosyakova, A. I. Buzin, V. G. Shevchenko, Yu. N. Luponosov, S. A. Ponomarenko, *Polymer Science, Series C* **2022**, *64*, 200.

[38]   X. Chen, X. Han, Q. Shen, *Adv Electron Mater* **2017**, *3*, DOI 10.1002/aelm.201600460.

[39]   E. Jia, D. Wei, P. Cui, J. Ji, H. Huang, H. Jiang, S. Dou, M. Li, C. Zhou, W. Wang, *Advanced Science* **2019**, *6*, DOI 10.1002/advs.201900252.

[40]   F. Cao, W. Tian, L. Meng, M. Wang, L. Li, *Adv Funct Mater* **2019**, *29*, DOI 10.1002/adfm.201808415.

[41]   S. Wang, Q. Li, *IET Nanodielectrics* **2018**, *1*, 80.

[42]   T. Gallet, D. Grabowski, T. Kirchartz, A. Redinger, *Nanoscale* **2019**, *11*, 16828.

[43]   K. Sotthewes, R. Van Bremen, E. Dollekamp, T. Boulogne, K. Nowakowski, D. Kas, H. J. W. Zandvliet, P. Bampoulis, *Journal of Physical Chemistry C* **2019**, *123*, 5411.

[44]   G. Alkhalifah, A. D. Marshall, F. Rudayni, S. Wanigasekara, J. Z. Wu, W.-L. Chan, *J Phys Chem Lett* **2022**, *13*, 6711.

[45]   A. S. Shikoh, A. Y. Polyakov, P. Gostishchev, D. S. Saranin, I. V. Shchemerov, S. I. Didenko, A. Di Carlo, *Appl Phys Lett* **2021**, *118*, DOI 10.1063/5.0037776.

[46]   D. Li, J.-Y. Shao, Y. Li, Y. Li, L.-Y. Deng, Y.-W. Zhong, Q. Meng, *Chemical Communications* **2018**, *54*, 1651.

[47]   A. Al-Ashouri, E. Köhnen, B. Li, A. Magomedov, H. Hempel, P. Caprioglio, J. A. Márquez, A. B. Morales Vilches, E. Kasparavicius, J. A. Smith, N. Phung, D. Menzel, M. Grischek, L. Kegelmann, D. Skroblin, C. Gollwitzer, T. Malinauskas, M. Jošt, G. Matič, B. Rech, R. Schlatmann, M. Topič, L. Korte, A. Abate, B. Stannowski, D. Neher, M. Stolterfoht, T. Unold, V. Getautis, S. Albrecht, *Science (1979)* **2020**, *370*, 1300.

[48]   Z. Zhu, Y. Bai, T. Zhang, Z. Liu, X. Long, Z. Wei, Z. Wang, L. Zhang, J. Wang, F. Yan, S. Yang, *Angewandte Chemie International Edition* **2014**, *53*, 12571.

[49]   W. Chen, Y. Wu, J. Fan, A. B. Djurišić, F. Liu, H. W. Tam, A. Ng, C. Surya, W. K. Chan, D. Wang, Z. He, *Adv Energy Mater* **2018**, *8*, DOI 10.1002/aenm.201703519.

[50]   F. Cao, W. Tian, L. Meng, M. Wang, L. Li, *Adv Funct Mater* **2019**, *29*, DOI 10.1002/adfm.201808415.

[51]   E. Jia, D. Wei, P. Cui, J. Ji, H. Huang, H. Jiang, S. Dou, M. Li, C. Zhou, W. Wang, *Advanced Science* **2019**, *6*, DOI 10.1002/advs.201900252.

[52]   S. Zeiske, W. Li, P. Meredith, A. Armin, O. J. Sandberg, *Cell Rep Phys Sci* **2022**, *3*, 101096.

[53]   P. Caprioglio, M. Stolterfoht, C. M. Wolff, T. Unold, B. Rech, S. Albrecht, D. Neher, *Adv Energy Mater* **2019**, DOI 10.1002/aenm.201901631.

[54]  J. Bisquert, C. Gonzales, A. Guerrero, *The Journal of Physical Chemistry C* **2023**, *127*, 21338.

[55]  E. Ghahremanirad, A. Bou, S. Olyaee, J. Bisquert, *J Phys Chem Lett* **2017**, *8*, 1402.

[56]  K. Sakhatskyi, R. A. John, A. Guerrero, S. Tsarev, S. Sabisch, T. Das, G. J. Matt, S. Yakunin, I. Cherniukh, M. Kotyrba, Y. Berezovska, M. I. Bodnarchuk, S. Chakraborty, J. Bisquert, M. V. Kovalenko, *ACS Energy Lett* **2022**, *7*, 3401.

[57]  A. Mahapatra, V. Anilkumar, R. D. Chavan, P. Yadav, D. Prochowicz, *ACS Photonics* **2023**, *10*, 1424.

[58]  J. Ding, W. Gao, L. Gao, K. Lu, Y. Liu, J.-L. Sun, Q. Yan, *J Phys Chem Lett* **2022**, *13*, 7831.

[59]  K. Kato, S. Hata, J. Yoshida, A. Kozen, in *LEOS 1992 Summer Topical Meeting Digest on Broadband Analog and Digital Optoelectronics, Optical Multiple Access Networks, Integrated Optoelectronics, and Smart Pixels*, IEEE, **n.d.**, pp. 254–257.

[60]  A. Armin, M. Hambsch, I. K. Kim, P. L. Burn, P. Meredith, E. B. Namdas, *Laser Photon Rev* **2014**, *8*, 924.

[61]  L. Shen, Y. Fang, D. Wang, Y. Bai, Y. Deng, M. Wang, Y. Lu, J. Huang, *Advanced Materials* **2016**, *28*, 10794.

[62]  F. Cao, W. Tian, L. Meng, M. Wang, L. Li, *Adv Funct Mater* **2019**, *29*, DOI 10.1002/adfm.201808415.

[63]  Y. Aggarwal, J. Park, W. Kim, S. Sharma, H. Jeong, M. G. Kim, J. Kil, E. H. Choi, B. Park, *Solar Energy Materials and Solar Cells* **2024**, *270*, 112815.

[64]  W. Kim, J. Park, Y. Aggarwal, S. Sharma, E. H. Choi, B. Park, *Nanomaterials* **2023**, *13*, 619.

# The supplementary information for the paper:

## Improvement of the perovskite photodiodes performance via advanced interface engineering with polymer dielectric

A.P. Morozov[1,§], L.O. Luchnikov[1,§], S. Yu. Yurchuk[2], A.R. Ishteev[1], P.A. Gostishchev[1], S.I. Didenko[2], N.S. Saratovsky[3], S.S. Kozlov[4], D.S. Muratov[5*], Yu. N. Luponosov[3*] and D.S. Saranin[1*]

[1]LASE – Laboratory of Advanced Solar Energy, NUST MISiS, 119049 Moscow, Russia
[2]Department of semiconductor electronics and device physics, NUST MISiS, 119049 Moscow, Russia
[3]Enikolopov Institute of Synthetic Polymeric Materials of the Russian Academy of Sciences (ISPM RAS), Profsoyuznaya St. 70, Moscow, 117393, Russia

[4]Laboratory of Solar Photoconverters, Emanuel Institute of Biochemical Physics, Russian Academy of Sciences, 119334 Moscow, Russia

[5]Department of Chemistry, University of Turin, 10125, Turin, Italy

**Experimental section:**

*Materials*

All organic solvents—dimethylformamide (DMF), N-Methylpyrrolidone (NMP), isopropyl alcohol (IPA), chlorobenzene (CB) were purchased in anhydrous, ultra-pure grade from Sigma Aldrich, and used as received. P(VDF-TrFE) (70:30 mol%) was purchased from Sigma Aldrich. Ethylacetate (EAC, 99+% purity) was purchased from Reaktivtorg-Himprocess hps, 2-Methoxyethanol was purchased from Acros Organics (99.5+%, for analysis), $HNO_3$ (70%). Photodiodes were fabricated on $In_2O_3$: $SnO_2$ (ITO) coated glass ($R_{sheet}$<7 Ohm/sq) from Zhuhai Kaivo company (China). $NiCl_2·6H_2O$ (from ReaktivTorg 99+% purity) used for HTM fabrication. Lead Iodide (99.9%), Cesium iodide (99.99%), Cesium chloride (99.99%) trace metals basis from LLC Lanhit, Russia and formamidinium iodide (FAI, 99.99% purity from GreatcellSolar), were used for perovskite ink. [6,6]-Phenyl-C61-butyric acid methyl ester (99% purity) was purchased from MST NANO (Russia). Bathocuproine (BCP, >99.8% sublimed grade) was purchased from Osilla Inc. (UK) and used for the fabrication of hole blocking layer.

*Inks preparation*

For the preparation of composition $Cs_{0.2}FA_{0.8}PbI_{2.93}Cl_{0.07}$ perovskite ink, we used CsCl, CsI, FAI, $PbI_2$ powders in a 0.07:0.13:0.8:1 molar ratio. The resulting mixture was dissolved in a DMF:NMP (volume ratio 640:360) with a concentration of 1.35 M and stirred at a temperature of 50 °C for 1 h. PCBM was dissolved in CB at a concentration of 27 mg/ml and stirred for 1 h at a temperature of 50 °C. BCP was dissolved in IPA at a concentration of 0.5 mg/ml and stirred for 8 h at a temperature of 50 °C. Before use, all solutions were filtered through 0.45 μm PTFE filters. P(VDF-TrFE) was dissolved in a perovskite solution at a concentration of $5 \times 10^{-3}$ mg/ml. For bulk/n-side configuration P(VDF-TrFE) was dissolved in EAC with DMF (volume ratio 200:1) at a concentration of $5 \times 10^{-3}$ mg/ml and stirred for 1 h at a temperature of 50 °C

*Device fabrication*

Perovskite photodiodes were fabricated with inverted planar architecture ITO/c-NiO /perovskite ($Cs_{0.2}FA_{0.8}PbI_{2.93}Cl_{0.07}$)/PCBM/BCP/Cu. Firstly, the patterned ITO substrates were cleaned with detergent, de-ionized water, acetone, and IPA in the ultrasonic bath. Then, substrates were activated under UV-ozone irradiation for 30 min. $NiCl_2·6H_2O$ precursor for NiO HTM film was spin-coated at 4000 RPMs (30 s), dried at 120 °C (10 min), and annealed at 300 °C (1 h) in the ambient atmosphere. Perovskite absorber film was crystallized on the top of HTM with solvent engineering method. Perovskite precursor was spin-coated at 3000 RPMs (5 s), and 5000 RPMs (30 s), 200 μL of EAC were poured on the substrate on the 21st second after the start of the rotation process. Then, substrates were annealed at 85 °C (1 min) and 105 °C (30 min) for conversation into the black perovskite phase. Deposition process of perovskite with P(VDF-TrFE) addition was the same. For bulk/n-side configuration PVDF was spin-coated upon annealed perovskite film at 4000 RPMs (30s) and annealed at 105 °C (5 min). The PCBM ETL was spin-coated at 4000 RPMs

(30 s) and annealed at 50 °C (5 min). BCP interlayer was also spin-coated at 4000 RPMs (30 s) and annealed at 50 °C (5 min). The copper cathode was deposited with the thermal evaporation method at 2 × 10⁻⁶ Torr vacuum level. All devices were encapsulated with UV epoxy from Osilla inc. UV LS processes (P1-P3) were described in the manuscript.

*Laser scriber*

The laser scriber system was designed by LLC Nordlase (Russia).

Laser type – Nd:YVO$_4$, 355 nm, impulse – 22 ns at 50kHz. Maximum power – 3W.

The positioning of the samples was realized using motorized XY stage from Standa (1 um resolution in XY movement).

The maximum attenuation of the system – 99%.

During scribing all substrates were fixed with vacuum chuck.

*Laser patterning cycle*

ITO scribing (P1) was performed using 3 W power at a rate of 5 mm/s (50 kHz, 1 pulse per 3 microns). Electrical isolation between the anode electrodes of the ITO for each pixel in the row was achieved by sequentially conducting 9 passes of the laser beam (50-micron diameter) with an offset of 10 microns.

The P2 process was realized in three passes (5 μm offset) at 1 W power at a speed of 5 mm/s. The width of the scribing line of the P2 process was ~60 μm. After deposition of the metal electrode in a vacuum, the P3 process required the removal of conductive material from the insulating zones between the ITO anode electrodes. Additionally, transverse scribing of the metal electrode was performed to form the final pixel geometry.

The pixel formation, indicated by the transverse metal scribing line in Fig. 2(a) in the manuscript, required the process to be performed at 1 W power, a speed of 2 mm/s, and three passes (10 μm offset). The width of the cut was 40 μm. Electrical isolation between pixels (P3 process) was achieved by ablating the metal contact using the P1 path (side isolation lines according to the top view in Fig. 2(a)). To ensure isolation, scribing was performed at a power of 3 W, speed of 20 mm/s for 1 pass of the laser beam with a cutting width of 40 μm. The P1-P3 patterning processes were the same for PPDs of all configurations.

*Characterization*

Surface roughness and film thicknesses were measured with KLA-Telencor stylus profilometer.

The fluorescence was induced with 375 nm picosecond pulsed laser (CNILaser MDL-PS-375). The signal acquisition was conducted until 15000 counts.

$$\alpha(h\nu) = \frac{B(h\nu - E_g)^{\frac{1}{2}}}{h\nu} \tag{S1}$$

Where $\alpha$ - absorption coefficient;

$h$ – Plank's constant;

$\nu$ – frequency;

$E_g$ – band gap.

**KPFM** measurements were conducted using an NT-MDT Ntegra AFM equipped with conductive probe NSG03/Au (NT-MDT). The work function was determined with a fresh HOPG surface serving as the energy baseline. The surface topography scans carried in tapping mode first and then a 1 V voltage was applied on the sample with tip resonance frequency (~200 kHz) to measure the sample surface potential ($V_{CPD}$) distribution through a DC voltage feedback loop. The scan rate was set to 0.8 Hz, and a lift scan height of approximately 50 nm was adopted. The work function ($W_f$) was calculated according to S1 and S2.

$$W_f^{tip} = W_f^{HOPG} - V_{CPD}^{HOPG} \qquad (S2)$$

$$W_f^{sample} = W_f^{tip} - V_{CPD}^{sample} \qquad (S3)$$

X-ray diffraction (**XRD**) patterns was measured with X-ray diffractometer Tongda TDM-10 using CuKα as a source with wavelength 1.5409 Å under 30 kV voltage and a current of 20 mA.

Steady state photoluminescence (**PL**) measurements were performed on Cary EclipseFluorescence Spectrophotometer with an excitation wavelength of 550 nm. Absorption spectra were measured on Spectronic HeliosAlpha UV–vis spectrophotometer from Thermo Fisher Scientific (Waltham, MA, USA). The Tauc plots calculation was realized using equation:

The dark JV-curves measurements were performed in the dark box in an ambient atmosphere with Keithley 2400 SMU in 4-wire mode (voltage step of 20 mV).

We used TDS-P003L4F07 (540 nm) as a light source for estimation of $V_{oc}$, $J_{sc}$ vs. $P_0$. The output parameters ($V_{oc}$, $J_{sc}$) were extracted from JV curves measured with Keithley 2400 SMU in 4-wire mode and a settling time of $10^{-2}$ s. The LED was connected to a GW Instek PSP-603 source in a hinge-mounted configuration. Optical power measurements were performed on a ThorLabs S425C. Illuminance measurements were performed on a UPRtek MK350. Completed set up for characterization of PPDs was placed in a black box.

The dynamic response and $f_{3dB}$ was measured with Tektronix TDS 3054C (oscilloscope) and Tektronix AFG 3252 (pulse generator). We used TDS-P001L4G05 STAR LED (540 nm) as a light source. Completed set up for characterization of PPDs was placed in a black box.

The **PPD**s stability was assessed by the change in response speed after heating for 250 hours at a temperature of 70 °C. Data was obtained for 30 PPDs of each configuration.

The **EQE** spectra were measured using QEX10 solar cell quantum efficiency measurement system (PV Measurements Inc., USA) equipped with xenon arc lamp source and dual grating monochromator. Measurements were performed in DC mode in the 300–850 nm range at 10 nm step. The system was calibrated using the reference NIST traceable Si photodiode.

Time resolved photoluminescence **(TRPL)** measurements was performed with time correlated single photon counter technique (TCSPC) on Zolix OmniFluo-990 spectrofluorometer.

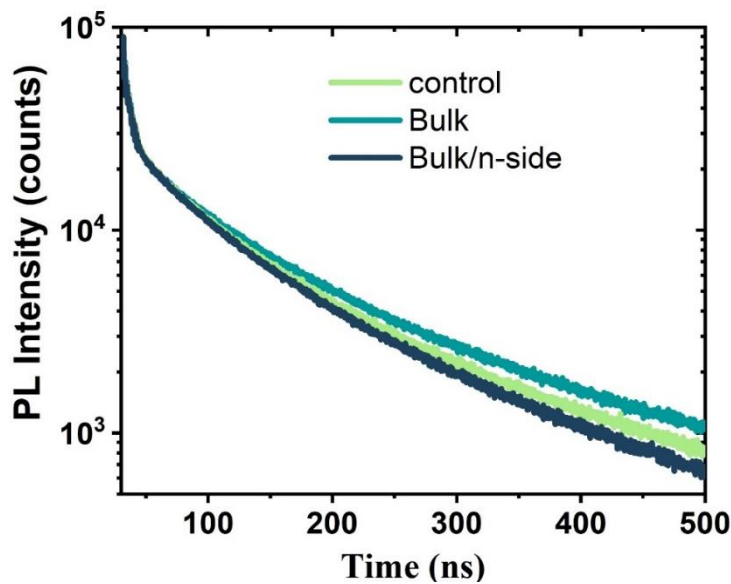

Figure S1 – The TRPL spectra of the PPDs with dielectric interlayers

**Details for the double diode model used for fitting of dark JV curves**

The equation for the diode current-voltage curve, including series (Rs) and shunt (Rsh) resistance, has the following expression (S1):

$$J = J_0 \cdot \left( \exp\left( \frac{q(V - J \cdot R_s)}{m \cdot k \cdot T} \right) - 1 \right) + \frac{V - J \cdot R_s}{R_{sh}}. \tag{S4}$$

For common pn junction, such a model describes the characteristics of real structures quite accurately. However, for a p-i-n structure, such a model does not always allow us to express the characteristic corresponding to experimental results. The reason for this is the presence of two barriers from the p- and n-regions, therefore, to calculate the current-voltage characteristics of the p-i-n structure, we used a two-diode model, represented by the equivalent circuit in Fig. S2. [1]

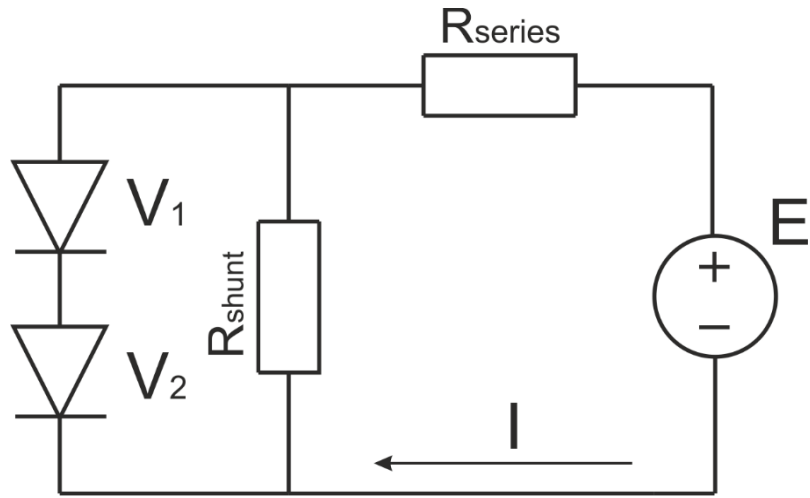

**Figure S2** – Double diode circuit for modeling of pin PSC

Diodes 1 and 2 in this circuit are ideal diodes, the current-voltage characteristics of which are described by the expressions S2-S3:

$$J_1 = J_{01} \cdot \left( \exp\left( \frac{qV_1}{m_1 \cdot k \cdot T} \right) - 1 \right) \tag{S5}$$

$$J_2 = J_{02} \cdot \left( \exp\left( \frac{qV_2}{m_2 \cdot k \cdot T} \right) - 1 \right) \tag{S6}$$

Two diode structures are connected in series, so the currents are equal to each other.

$$J_d = J_{01} \cdot \left( \exp\left( \frac{qV_1}{m_1 \cdot k \cdot T} \right) - 1 \right) = J_{02} \cdot \left( \exp\left( \frac{qV_2}{m_2 \cdot k \cdot T} \right) - 1 \right) \tag{S7}$$

Since the equivalent circuit is branched, it is necessary to solve a system of equations obtained from Kirchhoff's laws to calculate the current-voltage characteristic, For a given voltage V, unknown values are the currents in the diode and shunt resistance circuits $J_d$ and Rsh. The voltages applied to each diode $V_1$ and V2, and the total current J, which we must find for each given voltage. The system of equations (S4) - (S7) is sufficient for the numerical calculation of the current – voltage characteristics according to the two-diode model, but it is necessary to calculate the model parameters: leakage currents of each individual diode $J_{01}$ and $J_{02}$, non-ideality coefficients $m_1$ and $m_2$ of each diode, series resistance $R_s$ and shunt resistance $R_{sh}$.

$$V_1 + V_2 = V - J \cdot R_s, \tag{S8}$$

$$J_{R_{sh}} = \frac{V - J \cdot R_s}{R_{sh}}, \tag{S9}$$

$$J = J_d + J_{R_{sh}}. \tag{S10}$$

The calculation was done with one of the methods for multi-parameter optimization, in which the objective function requiring minimization of the sum for the differences between the experimental and theoretically calculated currents (S8).

$$\sum_{i=1}^{m}(J_{ex_i} - J_{teor_i})^2, \tag{S11}$$

where m is the number of experimental points.

Calculation program was developed in Borland Delphi 7, which allows determining the parameters of a two-diode structure using multi-parameter optimization by the coordinate descent method.

$$R = \frac{I_{ph}}{P_0} \tag{S12}$$

Where $I_{ph}$- photocurrent (A)

$P_0$- power of the illumination (W)

$$LDR = 20 \log\left(\frac{I_{ph}}{I_D}\right) \tag{S13}$$

Where $I_{ph}$- photocurrent in linear range

$I_D$- dark current

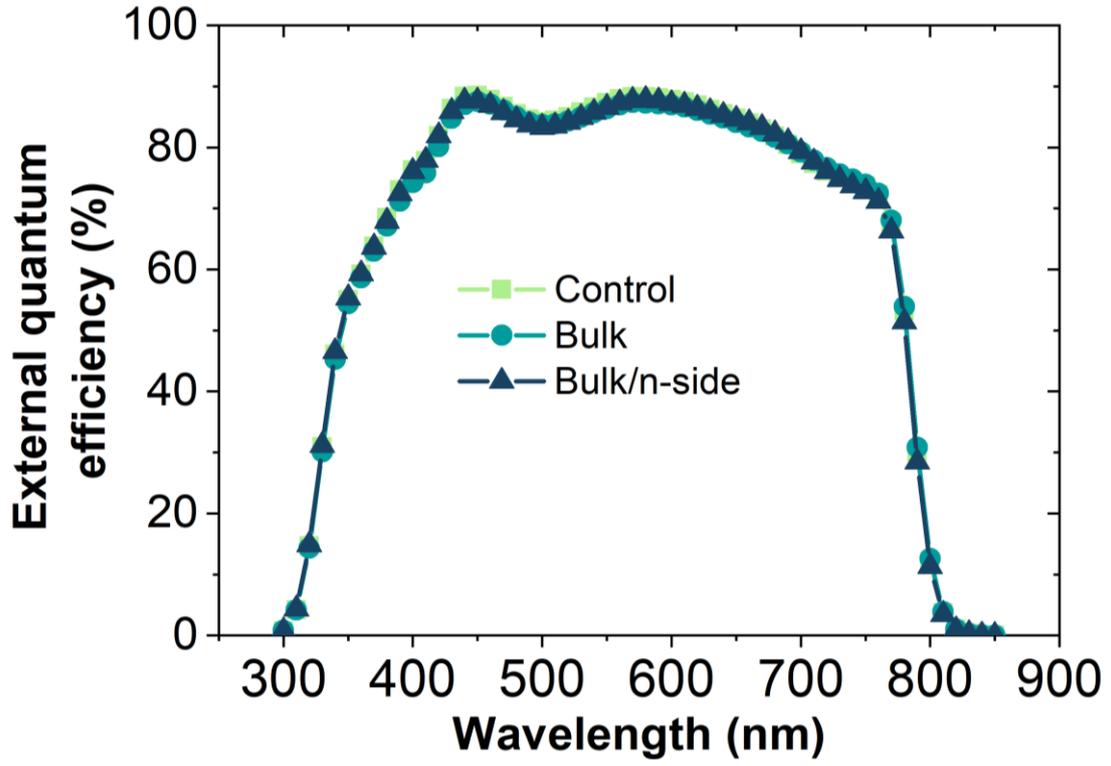

Figure S3 – The external quantum efficiency of the PPDs with dielectric interlayers

$$D^* = \frac{I_{ph}\sqrt{A}}{P_0}\left(\frac{1}{(2qJ_d)^{1/2}}\right) \tag{S14}$$

Where A – active area of the device

$$NEP = \frac{(2qJ_d)^{1/2}}{R} \tag{S15}$$

$$|I_{ph}| = \eta_{abs}\eta_{qe}\eta_{ce}\frac{q\lambda}{hc}I_L \tag{S16}(Zeiske\ et\ al.,\ 2022)$$

Where $I_{ph}$- photocurrent;

$\eta_{abs}$- absorptance;

$\eta_{abs}$- quantum efficiency for the photogenerated charge carriers;

$\eta_{ce}$ -charge collection efficiency;

q – elementary charge;

λ- wavelength;

h-Planck's constant;

c – speed of light;

$I_L$- light intensity.

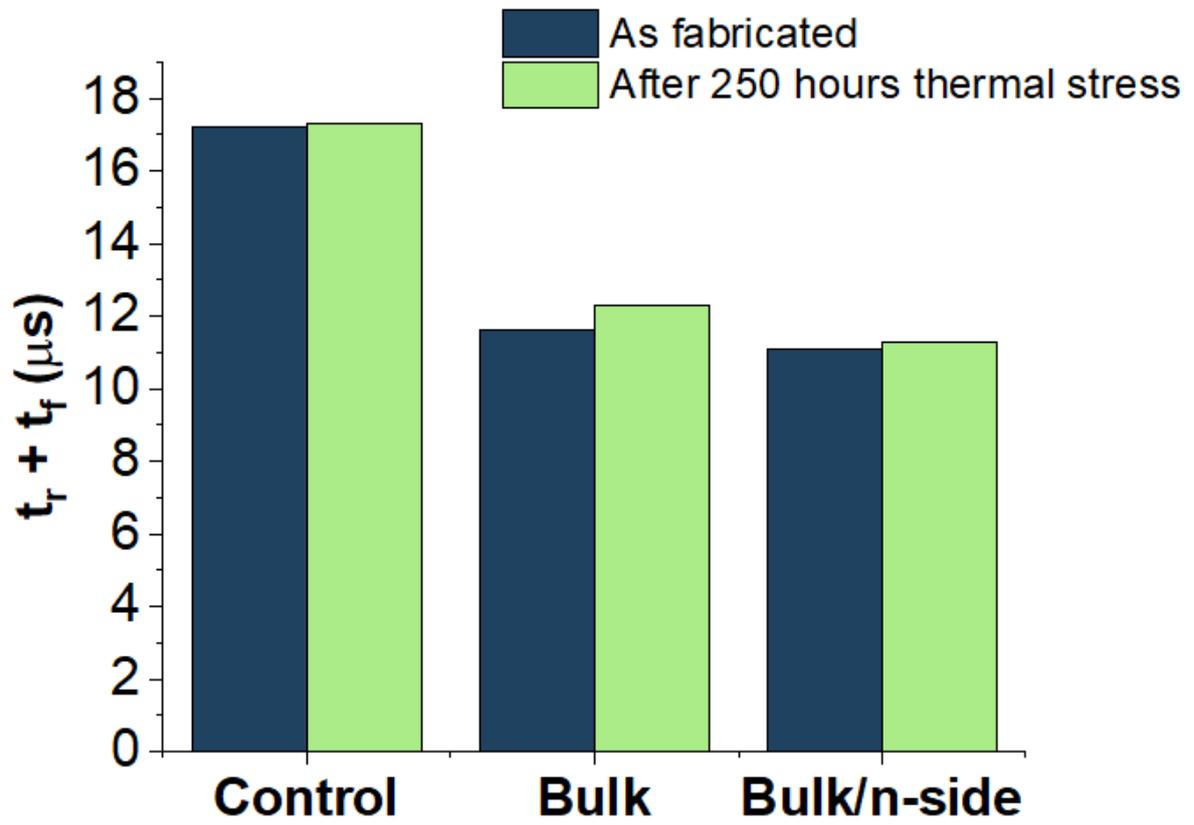

Figure S4 – Response speed of the devices before and after thermal stress

**References:**

[1] P. Liao, X. Zhao, G. Li, Y. Shen, M. Wang, *Nano-Micro Letters* **2018**, DOI 10.1007/s40820-017-0159-z.

[2] S. Zeiske, W. Li, P. Meredith, A. Armin, O. J. Sandberg, *Cell Rep Phys Sci* **2022**, *3*, 101096.